\documentclass[a4paper,useAMS,usenatbib]{mn2e}
\usepackage{graphicx, amsmath,mathcomp,amssymb,txfonts}
\usepackage{hyperref}

\title[Global Gravitational Instabilities in Discs with Infall]{Global Gravitational Instabilities in Discs with Infall}
\author[D.~Harsono, R. D. Alexander and Y. Levin]{D. Harsono$^1$\thanks{E-mail:
harsono@strw.leidenuniv.nl}, R. D. Alexander$^{1,2}$ and Y. Levin$^{1,3,4}$\\
$^1$Sterrewacht Leiden, Universiteit Leiden, Niels Bohrweg 2, 2300 RA, Leiden, the Netherlands\\
$^2$Department of Physics \& Astronomy, University of Leicester, Leicester, LE1 7RH\\
$^3$Lorentz Institute, P.O. Box 9506, 2300 RA Leiden, The Netherlands\\
$^4$School of Physics, Monash University, P.O.Box 27, VIC 3800, Australia
}

\begin{document}

\pagerange{\pageref{firstpage}--\pageref{lastpage}} \pubyear{2010}

\maketitle

\label{firstpage}

\begin{abstract}

Gravitational instability plays an important role in driving gas accretion in massive protostellar discs. Particularly strong is the global gravitational instability, which arises when the disc mass is of order 0.1 of the mass of the central star and has a characteristic spatial scale much greater than the disc's vertical scale-height. In this paper we use three-dimensional numerical hydrodynamics to study the development of gravitational instabilities in a disc which is embedded in a dense, gaseous envelope.  We find that global gravitational instabilities are the dominant mode of angular momentum transport in the disc with infall, in contrast to otherwise identical isolated discs.  The accretion torques created by low-order, global modes of the gravitational instability in a disc subject to infall are larger by a factor of several than an isolated disc of the same mass.  We show that this global gravitational instability is driven by the strong vertical shear at the interface between the disc and the envelope, and suggest that this process may be an important means of driving accretion on to young stars.

\end{abstract}


\section{Introduction}\label{sec:intro}

Accretion discs play a fundamental role in many aspects of astrophysics.  Objects as diverse as planets, stars and super-massive black holes are all thought to acquire significant fractions of their mass through disc accretion, and consequently understanding accretion disc physics is important in understanding the formation of all of these objects.  Critical to our understanding of accretion discs is the process of angular momentum transport, but despite many years of research on this subject we still do not fully understand the mechanism(s) by which angular momentum is transported in gaseous discs.  In many cases we believe that magnetohydrodynamic instabilities, such as the magnetorotational instability \citep[MRI,][]{bh91,bh98} is the dominant transport mechanism.  However, some systems, notably protostellar discs, are insufficiently ionized for the MRI to operate everywhere \citep[e.g.,][]{g96}, and it is also not clear whether or not the MRI can drive accretion rates as high as those which are observed \citep*{kpl07}.  Consequently, it is still desirable to investigate other mechanisms for angular momentum transport in discs.

\begin{figure}
\centering
\resizebox{\hsize}{!}{
\includegraphics[angle=270]{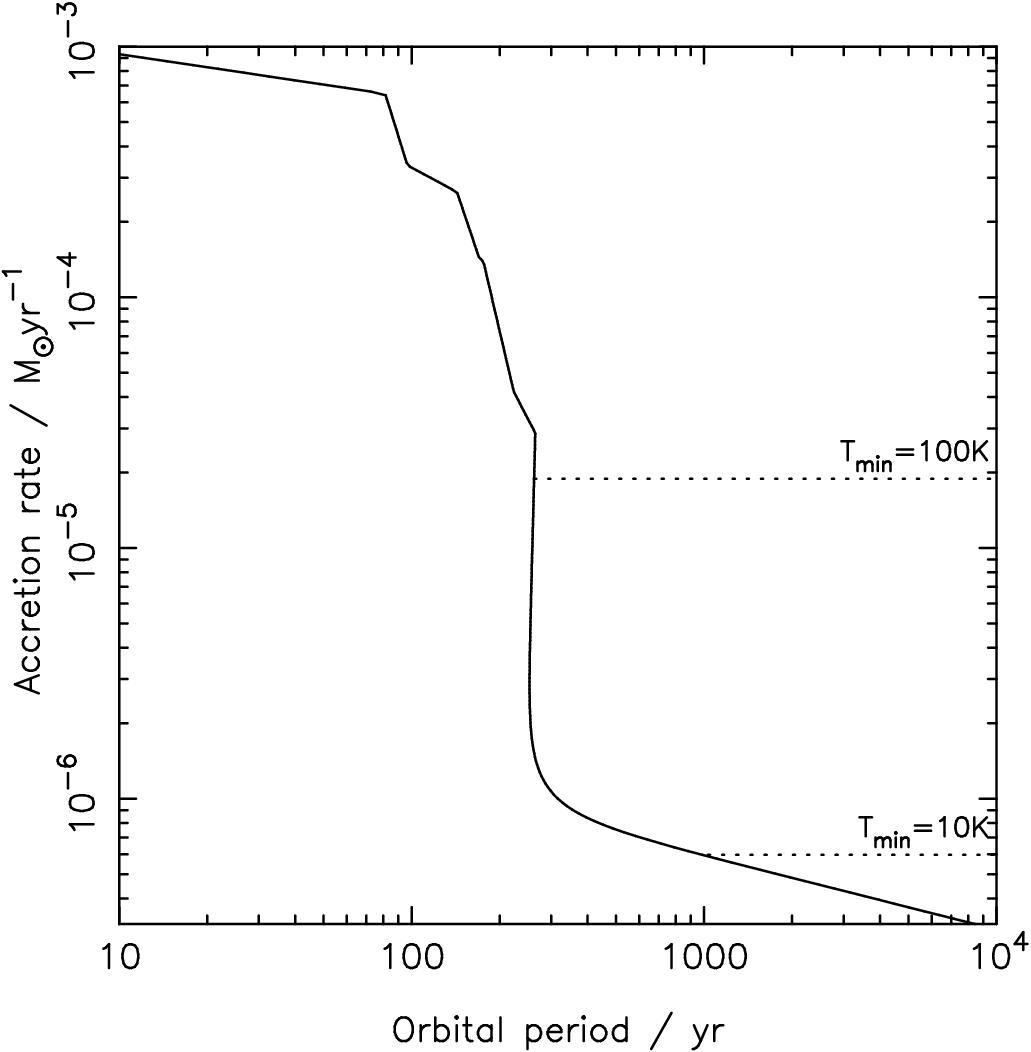}
}
\caption{Maximum sustainable accretion rate in a critically self-gravitating disc with $\alpha = 0.1$, computed following the procedure described in \citet{levin07} and using the opacities $\kappa(\rho,T)$ of \citet{bl94} and \citet{bell97}.  The sharp jump in the critical accretion rate at an orbital period of $\simeq 300$yr is caused by the transition between the optically thick inner disc and optically thin outer disc \citep{ml05}.  This corresponds to a radius of $\simeq 40$AU for a 1M$_{\odot}$ central star, or $\simeq 100$AU for a 10M$_{\odot}$ star.  The maximum sustainable accretion rate at larger radii is small, $\sim 10^{-6}$M$_{\odot}$yr$^{-1}$; in the ``local limit'', larger accretion rates lead to fragmentation.  In some cases external irradiation can be the dominant source of heating, imposing a temperature ``floor'' (denoted by $T_{\mathrm {min}}$) and enhancing the maximum accretion rate.  Similar figures can be found in \citet{clarke09} and \citet{rafikov09}.}
\label{fig:mdot_max}
\end{figure}

One such mechanism which has received considerable interest in recent years is angular momentum transport by gravitational instabilities \citep[GIs; see, e.g.,][and references therein]{d07,lodato08}.  Gaseous discs in Keplerian rotation become unstable to self-gravity when the \citet{toomre64} $Q$ parameter is less than some critical value of order unity.  The Toomre parameter is defined as
\begin{equation}
\label{eq:toomreq}
Q = \frac{c_{s} \Omega}{\pi G \Sigma} \, ,
\end{equation}
 where $c_{s}$ is the sound speed of the gas, $\Omega$ is the angular frequency and $\Sigma$ is the surface density.  Shearing discs generally become unstable to non-axisymmetric perturbations before axisymmetric ones, so GIs in discs initially manifest themselves as spiral density waves.  It was recognised long ago that such spiral density waves can transport angular momentum \citep{lbk72} but detailed study of the non-linear development of GIs in gaseous discs has only recently become possible.  This process has been studied in great detail using numerical hydrodynamics, and we now have a well-established picture whereby angular momentum transport by GIs is primarily governed by disc thermodynamics.  GIs in isolated thin gas discs tend to evolve to a self-regulating state, where the energy liberated by accretion is balanced by local (radiative) cooling \citep*[e.g.,][]{g01,lr04,mejia05,clc09}, and although gravity is a long-range force, ``global'' effects generally do not dominate unless the disc mass is an appreciable fraction ($\gtrsim 25$\%) of the mass of the central object \citep*{lka98,lr05}.  In this picture the efficiency of angular momentum transport can be parametrized in terms of a classical \citet{ss73} $\alpha$-prescription \citep{g01,lr04}, where
\begin{equation}
\alpha_{\mathrm {GI}} = \frac{4}{9}\frac{1}{\gamma (\gamma-1)t_{\mathrm {cool}}\Omega} \, .
\end{equation}
Here $t_{\mathrm {cool}}$ is the local cooling time-scale and $\gamma$ is the adiabatic index of the gas.  Faster cooling leads to deeper spiral density waves (i.e., with higher density contrasts), and thus to more efficient transport of angular momentum.  However, if the cooling becomes too rapid the disc is unable to maintain its self-regulating state, and the GIs instead lead to fragmentation of the disc \citep{g01,r03}.  This in turn imposes a maximum efficiency at which angular momentum can be transported by GIs without leading to disc fragmenting, and numerical simulations place typically this ``fragmentation boundary'' at $\alpha_{\mathrm {GI}} \lesssim 0.1$ \citep*[corresponding to $\beta = t_{\mathrm {cool}}\Omega \gtrsim 3$--5, e.g.,][]{g01,r03,rla05}.  When extended to consider discs with realistic opacities, these results imply a maximum accretion rate that can be sustained by GIs in a self-regulating state \citep[e.g.,][see also Fig.~\ref{fig:mdot_max}]{levin03,ml05,levin07,clarke09,rafikov09}.  Except at very small radii this rate is low, $\sim 10^{-6}$M$_{\odot}$yr$^{-1}$, and this raises questions as to how many astrophysical objects are able to accrete their mass in a plausible time-scale.  The star may continue accreting bound clumps of gas even after the disc fragments \citep[eg.,][]{vb10}, but the details of this process remain uncertain.

To date most numerical studies of GIs have looked at isolated self-gravitating discs, but in reality it seems likely that most gravitationally unstable discs will still be subject to some level of infall on to the disc.  Indeed, in many cases it is likely that the instantaneous infall rate on to the disc exceeds the accretion rate through the disc.  For example, observed accretion rates on to protostellar discs are typically an order of magnitude larger than the accretion rates on to the protostars themselves \citep*[e.g.,][]{kenyon90,calvet_ppiv}.  Similar discrepancies between infall and disc accretion rates have been found in models of low-mass star formation \citep[e.g.,][]{vorobyov09}, and in models of star formation in black hole accretion discs \citep[e.g.,][]{ml04}.  In this paper we present an initial investigation of this problem, by using three-dimensional numerical hydrodynamics to follow the evolution of a self-gravitating accretion disc subject to quasi-spherical infall.  In Section \ref{sec:method} we present our numerical method, and in Section \ref{sec:results} we discuss the results of our simulations.  We find that infall on to the disc can substantially enhance the efficiency of angular momentum transport, through the excitation of low-order, global, spiral density waves.  We discuss the consequences of this result for real astrophysical systems, along with the limitations of our analysis, in Section \ref{sec:dis}, and summarize our conclusions in Section \ref{sec:conc}.


\section{Numerical Method}\label{sec:method}

Our simulations are conducted using the publicly-available smoothed-particle hydrodynamic (SPH) code {\sc gadget-2} \citep{springel05}.  We have modified the code to include a simple scale-free cooling prescription, as used in previous simulations \citep{g01,lr04,clc09}, which has the following form:
\begin{equation}
\frac{d u_i}{dt} = - \frac{u_i} {t_{cool}} \, ,
\label{eq:cool}
\end{equation}
Here $u_i$ is the internal energy of particle $i$, and the cooling time-scale $t_{cool}$ is proportional to the local dynamical time-scale thus
\begin{equation}
t_{cool} = \frac{\beta}{\Omega} \, .
\label{eq:tcool}
\end{equation}
Operationally, the cooling time-scale is computed as
\begin{equation}
t_{cool} = \beta \sqrt{\frac{R_i^3}{G M_\star}} \, ,
\end{equation}
where $R_i$ is the cylindrical radius of the $i$th particle.  The cooling time thus depends only on radius, and does not vary with $z$ or with the instantaneous orbital speed (which can be perturbed significantly in unstable discs).  As mentioned in Section \ref{sec:intro}, previous simulations of self-gravitating discs have found that values of $\beta \lesssim 3$--5 result in fragmentation of the disc, while larger values lead to transport of angular momentum \citep[e.g.,][]{g01,r03,rla05}.  We do not wish to see disc fragmentation due to rapid cooling alone, and therefore set $\beta = 7.5$ throughout.  We adopt an adiabatic equation of state, with adiabatic index $\gamma = 5/3$.

We make use of a single sink particle as the central gravitating mass, which accretes all gas particles within its sink radius \citep[as described in][]{cuadra06}.  This is primarily a numerical convenience, used in order to prevent the time-step being limited by a small number of SPH particles at very small radii, and has no physical effect on the simulations.  We use the standard Barnes-Hut formalism to calculate the gravitational force tree, and use N$_{ngb} = 64 \pm 2$ as the number of SPH neighbours.  We allow a variable gravitational softening length, which is equal to the SPH smoothing length throughout \citep[as demanded by][]{nelson06}.  The simulations are scale-free: we use a system of units where the central gravitating mass has an initial mass $M_\star = 1$, the inner edge of the disc is at $R=1$\footnote{Note that we use upper-case $R$ to denote cylindrical radius, and lower-case $r$ for spherical radius.}, and the time unit is the orbital period at $R = 1$.  (Thus $G = 4\pi^{2}$ in code units.)

 \subsection{Artificial viscosity}
We adopt the standard Monaghan-Gingold-Balsara form for the artificial viscosity \citep{mg83,b95}, as described in Equations 11--12 of \citet{springel05}.  This prescription contains both linear and quadratic terms (characterised by the parameters $\alpha_{sph}$ and $\beta_{sph}$ respectively, with $\beta_{sph} = 2 \alpha_{sph}$), and the ``Balsara switch'' which acts to limit the artificial viscosity in pure shear flows.  We adopt $\alpha_{sph}=0.3$ throughout.

As we are primarily interested in how angular momentum is transported in our simulations, great care must be taken to ensure that this transport is not dominated by numerical effects.  It is well-known that SPH artificial viscosity can drive significant angular momentum transport in disc simulations \citep[e.g.,][]{m96,lr04}, so we have conducted tests to ensure that this is not the dominant source of transport in our models.  From our standard disc initial conditions (see Section \ref{sec:disc_ICs} below) we ran simulations with the self-gravity of the gas turned off; with this set-up, Reynolds stresses due to numerical effects (primarily the artificial viscosity) are the only source of angular momentum transport.  By expressing this stress in units of the local pressure \citep[see, e.g.,][]{lr04} we can parametrize the efficiency of the numerical transport as a familiar $\alpha$-parameter thus:
\begin{equation}
\alpha_{art} = \frac{2}{3} \frac{\delta v_r \delta v_{\phi}}{c_s^2} \, ,
\end{equation}
where $\delta {\mathbf v} = {\mathbf v} - \langle {\mathbf v} \rangle$ (i.e., the perturbation from the mean fluid velocity).  Except in the regions near the inner boundary ($R\lesssim 10$), where the flow is in any case dominated by boundary effects, we find that the efficiency of artificial transport is typically $\alpha_{art} \simeq 0.001$--0.005, and never exceeds 0.01.  Numerical transport of angular momentum is therefore at least an order of magnitude less efficient than the transport we expect from GIs, and we are confident that angular momentum transport by artificial viscosity does not have a strong influence on our results.

\subsection{Initial Conditions}\label{sec:ICs}
\subsubsection{Disc}\label{sec:disc_ICs}
Our discs are set up to use the same initial conditions as \citet{rla05}.  The central gravitating mass is surrounded by a gaseous disc with mass $M_d$, which is represented by 250,000 SPH particles.  The initial velocity profile is Keplerian, with a first-order (spherical) correction to account for the effect of the disc mass.  The disc extends from $R = 1$ to $R = 100$, with initial surface density and temperature profiles\begin{equation}
\Sigma (R) \propto R^{-1} \, ,
\label{densprof}
\end{equation}
and
\begin{equation}
T (R) \propto R^{-1/2} \, .
\label{temprofile}
\end{equation}
Thus $Q \propto R^{-3/4}$ (approximately), and we normalise the disc temperature so that $Q=2$ at the outer disc edge ($R=100$).  The disc is thus initially stable, and is allowed to cool into instability.  The vertical density distribution is Gaussian, with scale-height $H = c_s/\Omega$.  Because the disc's self-gravity is not negligible this configuration is not strictly in vertical hydrostatic equilibrium, but the discs adjust to equilibrium on a dynamical time-scale.  We performed two such simulations with different disc masses: $q = M_d/M_\star = 0.1$ and $q = 0.2$.  
\subsubsection{Spherical envelope}

\begin{figure}
\centering
\resizebox{\hsize}{!}{
\includegraphics[]{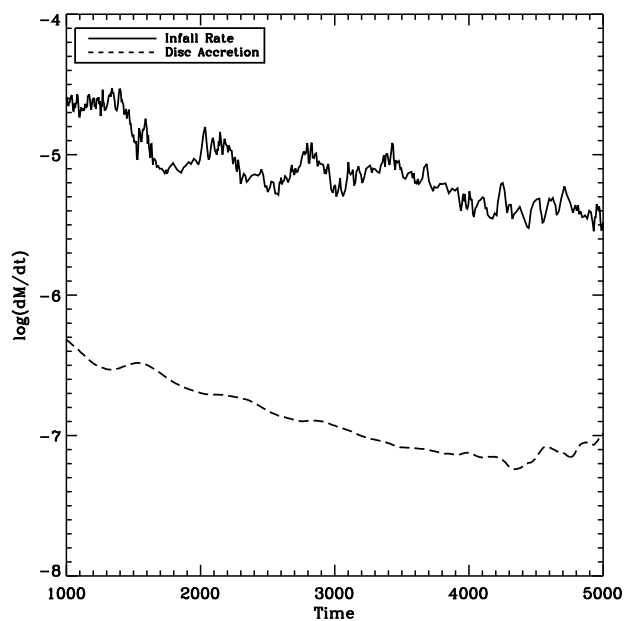}
}
\caption{Measured accretion rate on to the disc in the infall model, and the quasi-steady accretion rate through the $q = 0.1$ disc.  The quasi-steady rate is evaluated as $\dot{M} = 3 \pi \alpha_{\mathrm {GI}} c_s^2 \Sigma / \Omega$, with $\alpha_{\mathrm {GI}}=0.1$ (see text) and the sound speed evaluated at the disc midplane, averaged over the radial region $R=20$--80.  The infall rate is determined from the time derivative of the disc mass.}
\label{fig:infallrate}
\end{figure}

In order to study the effects of infall on the development of gravitational instabilities in the disc, we took the simplest possible approach and surrounded the $q=0.1$ disc with a uniform density spherical envelope.  The envelope has the same mass as the disc ($0.1M_\star$), and thus uses a further 250,000 SPH particles, and extends from $r=1$ to $r=500$.  The envelope is initially isothermal, with a temperature equal to that at the disc outer edge.  In order to prevent gas particles spiralling tightly around the vertical axis resulting in unreasonably short time-steps, the spherical envelope has a cylindrical hole around the $z$-axis which extends to $R = 10$.  The envelope was initially given solid body rotation, with the angular frequency fixed to be 0.08 of the Keplerian value at the outer disc radius.  This value was chosen so that the bulk of the envelope mass falls on to the disc away from the inner boundary, where the disc is numerically well-behaved.  With this set-up, most of the initial infall occurs at radii from $R\simeq20$--100. The measured infall rate is around an order of magnitude greater than the maximum quasi-steady accretion rate through the disc, as shown in Fig.~\ref{fig:infallrate}.  This discrepancy between the infall rate and the disc accretion rate can be understood as follows.  The infall is roughly spherical, and the cooling time-scale is long compared to the infall (dynamical) time-scale, so the infall rate is approximately
\begin{equation}
\dot{M}_{infall} \sim \frac{c_s^3}{G} \, .
\end{equation}
By contrast, the maximum sustainable accretion rate through the disc (in the local limit) is 
\begin{equation}
\dot{M}_{acc,max} \sim \alpha_{max} \frac{c_s^3}{G} \, .
\end{equation}
The sound speeds in these two equations are not necessarily the same: the first is in the envelope, while the second is in the disc midplane.  However, in our simulations the radial variation of $c_s$ in the disc is weak ($\propto R^{-1/4}$) and the cooling is slow, so in practice the two sound speeds are very similar.  The fragmentation boundary in isolated discs is $\alpha_{max} \simeq 0.1$ \citep{rla05}, and thus it is physically reasonable for $\dot{M}_{infall}$ to exceed $\dot{M}_{acc,max}$ by approximately a factor of 10.  This situation, however, naturally leads to an unsustainable bottleneck, as the infalling mass cannot be accreted in the ``normal'' manner.  We therefore expect this simulation to have dramatic results: presumably, the disc must either fragment (despite slow cooling), or undergo some sort of violent relaxation process (with rapid transport of angular momentum).  


\section{Results}\label{sec:results}
\begin{figure*}
\centering
\resizebox{\hsize}{!}{
\includegraphics[angle=270]{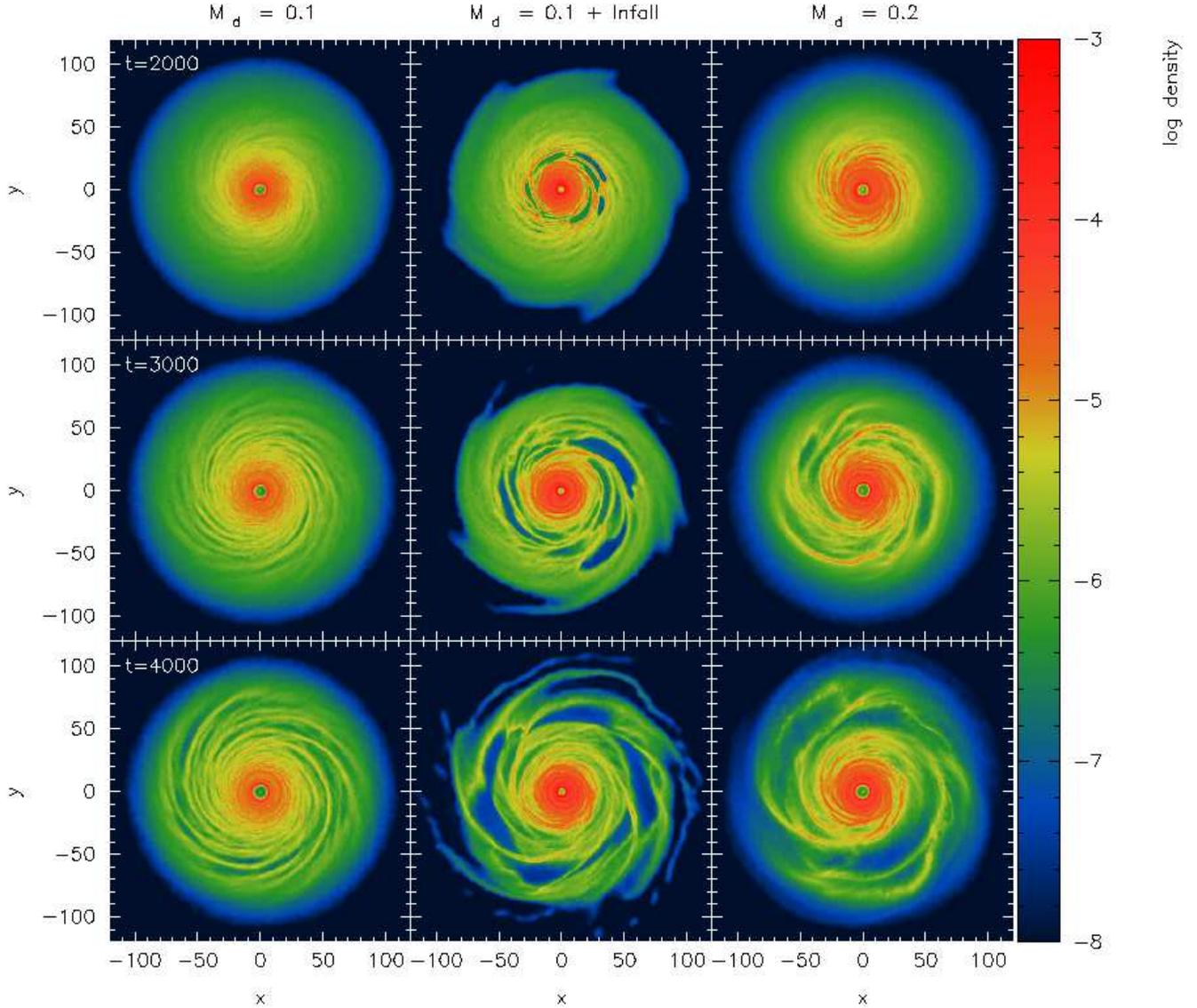}
}
\caption{Time evolution of midplane density in the different simulations.  From left-to-right, the three columns show the time evolution of the $q=0.1$, infall and $q=0.2$ discs respectively.  The isolated disc models evolve into a self-regulating state, with quasi-stable transport of angular momentum.  In the presence of infall, however, the disc shows dramatic departures from self-regulation, with high-amplitude spiral density waves, low-order spiral modes, and increased rates of angular momentum transport.}
\label{fig:panels}
\end{figure*}

We performed three simulations: two isolated disc simulations with $q = 0.1$ and $q=0.2$, and the ``infall'' (disc + spherical envelope) simulation.  The isolated disc simulations act as reference models: the $q=0.1$ disc is identical to that in the infall model (but with no infall), and the $q=0.2$ disc serves as a reference where all of the envelope mass instead initially resides in the disc.  The two isolated disc models essentially ``bracket'' the infall model (which has an initial mass of $0.1M_\star$ and a maximum final mass of $0.2M_\star$), and allow us to discriminate between mass and infall effects.  We followed each simulation to $t=5000$ (code units).  This corresponds to 5 outer disc orbital periods (i.e., slightly less than one outer cooling time-scale), after which time most of the material in the envelope has already fallen onto the disc\footnote{Note, however, that a significant fraction of the envelope mass falls to the midplane at $R>100$, beyond the outer edge of the initial disc.}.  In the region where the disc is numerically well-behaved (approximately $R=20$--80; away from the disc boundaries) we are thus able to follow the dynamics for tens of dynamical time-scales, allowing the instabilities to develop in a physical manner.

\subsection{Isolated Discs}\label{sec:disc_only}
Many previous studies have studied the transport properties of GI in low-mass ($q \lesssim 0.25$) discs \citep[e.g.,][]{lr04, boley06, clc09}.  Our low-mass ($q=0.1$) isolated disc is essentially identical to that used in previous simulations \citep{rla05,clc09}, and its expected behaviour is well understood.  The transport of angular momentum is dominated by high-order ($m\sim10$) spiral density waves, and the dynamics are consistent with the local approximation.  The characteristic length-scale of such spiral density waves is in the order of disc scale height $H$.  This simulation therefore serves two purposes here: it acts as a code test, as our calculation should reproduce previous results, and it also provides us with a reference model with which to compare our ``disc with infall'' calculation\footnote{For numerical reasons $q=0.1$ is also approximately the lowest mass disc that can be well-resolved in these simulations.  The resolution requirements in such simulations essentially amount to always ensuring that the disc scale-height is resolved into several SPH smoothing lengths \citep{nelson06}.  In a self-gravitating disc the scale-height is proportional to the disc mass, so for three-dimensional simulations a factor of two decrease in the disc mass typically costs more than an order of magnitude in computation time (an increase of $\sim 2^3$ in particle number, plus a shortening of the time-step).  Consequently, long-duration simulations of discs with $q \ll 0.1$ remain prohibitively expensive.}.  Previous studies have also shown that increasing the disc mass results in the GI generating more power in the lower-$m$, global spiral modes \citep{lr05}.  We therefore ran a second reference simulation, with $q=0.2$, so that we are able to distinguish the effects of infall from those that are simply due to the increasing disc mass.

\begin{figure}
\centering
\resizebox{\hsize}{!}{
\includegraphics[angle=270]{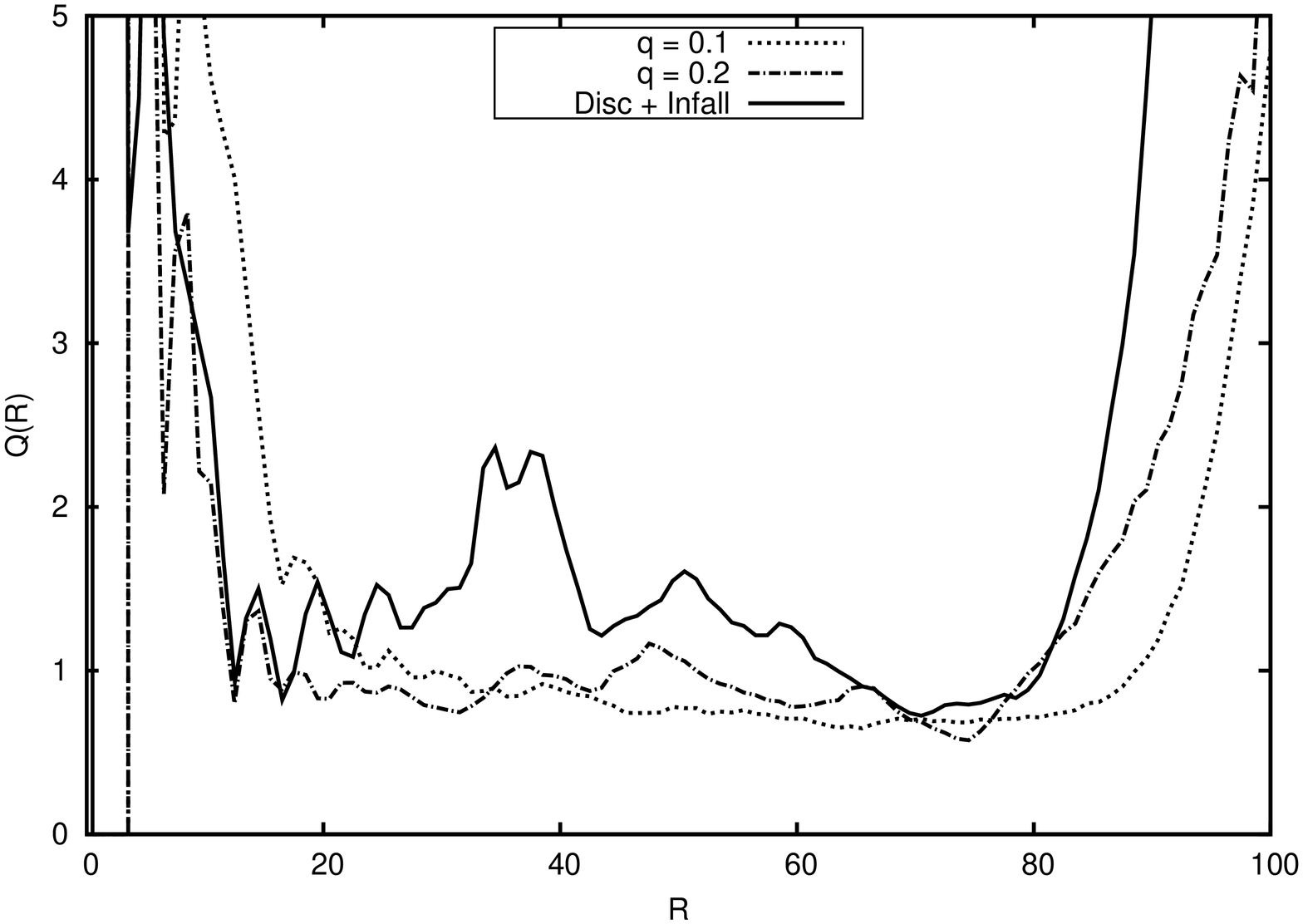}
}
\caption{Azimuthally-averaged Toome $Q$ parameter for the three different simulations, plotted at t = 3500.  Note the strong divergence from $Q\simeq1$ in the infall simulation.}
\label{fig:Qprof}
\end{figure}

Our isolated disc simulations essentially repeat these previous studies, and we observe the same general behaviour described in (for example) \citet{lr04} and \citet{clc09}.  The time evolution of the simulations is shown in Fig.~\ref{fig:panels}, which shows the midplane density evolution from $t = 2000$ to $t = 4000$.  The isolated discs initially cool and become gravitationally unstable, and then develop long-lived spiral density waves which transport angular momentum in a quasi-steady manner.  Both discs quickly settle into a self-regulating state, with $Q\simeq1$ at all radii (see Fig.~\ref{fig:Qprof}).  The maximum density contrast in the spiral arms reaches approximately 1.5 orders of magnitude.  The drop in surface density at small radii is due to the artificial pressure gradient introduced by the inner boundary, and is not physical; for this reason, we neglect the inner region of the discs ($R \le 20$) in our subsequent analysis.  Some additional power in low-order spiral modes is seen in the more massive ($q=0.2$) disc, but for the most part the transport is well-characterised by a local model, where energy released by accretion is locally balanced by the imposed cooling.

The induced global spiral density waves can be examined by decomposing the disc's structure into Fourier modes. In order to compute the Fourier amplitudes, we divided the disc into concentric annuli and computed the amplitudes of the azimuthal modes for each annulus.  We then integrated these amplitudes radially to give global Fourier amplitudes $A_m$, where $m$ is the Fourier mode:
\begin{equation}\label{eqn:azmamp}
A_{m} = \frac{1}{N_{disc}} \left | \sum_{R = 20}^{80} \sum_{j = 1}^{N_{ann}} \exp^{- i m \phi_j } \right | \, .
\end{equation}
Here $N_{ann}$ is the number of SPH particles in each annulus, and $N_{disc}$ is the total number of particles in all the annuli (i.e., the disc mass).  $m$ is the azimuthal mode number, and $\phi_j$ is the azimuthal angle (phase) of the $j$th particle.  We use annuli of width $\Delta R=1.0$, which gives $N_{ann} \sim 2000$ particles in each annulus.  Because of the strong influence the outer and (especially) inner boundaries, we limit ourselves to the radial range $20\le R \le80$; numerical effects are likely to be significant outside this range.

\begin{figure}
\centering
\resizebox{\hsize}{!}{
\includegraphics[angle=270]{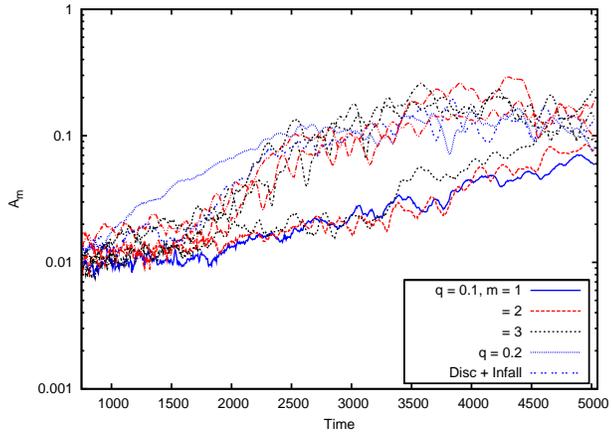}
}
\caption{Time evolution of the lowest-order Fourier modes, $A_{m}$, in the three simulations: q = 0.1 (dotted), q = 0.2 (dashed) and disc $+$ infall (solid).  Blue, red and black lines denote $m=1, 2$ \& 3 respectively.  The $q=0.1$ disc shows significantly less power in these low-order modes than the other models.}
\label{fig:mlow}
\end{figure}

\begin{figure}
\centering
\resizebox{\hsize}{!}{
\includegraphics[angle=270]{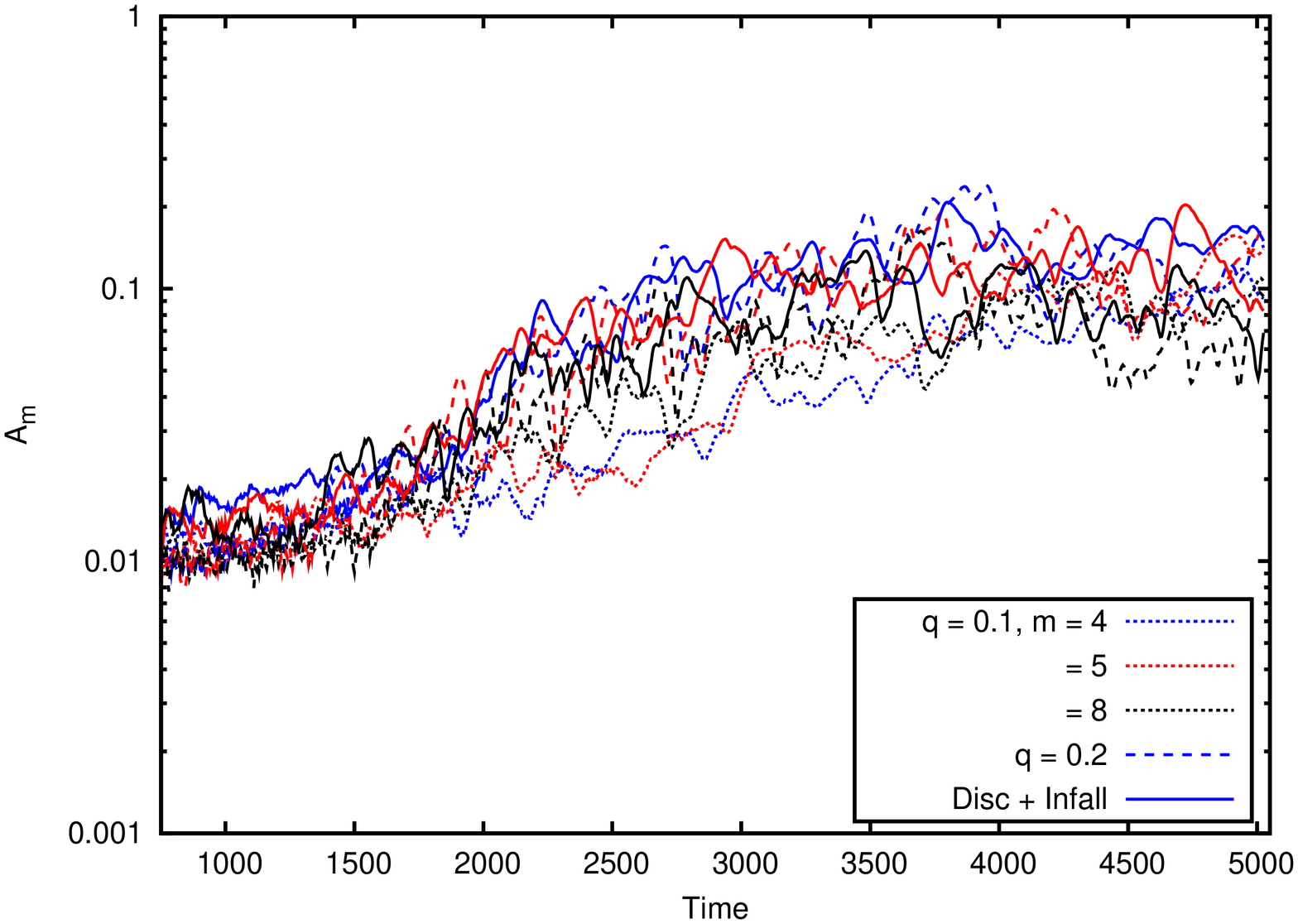}
}
\caption{Time evolution of higher-order Fourier modes, $A_{m}$, in the three simulations: q = 0.1 (dotted), q = 0.2 (dashed) and disc $+$ infall (solid).  Here blue, red and black denote $m=4, 5$ \& 8 respectively.  For these higher-order modes, the differences between the different models are much less pronounced than in Fig.\ref{fig:mlow}.}
\label{fig:mhigh}
\end{figure}

Figs.~\ref{fig:mlow} \& \ref{fig:mhigh} show the time evolution of the Fourier amplitudes in our simulations.  In the $q=0.1$ disc modes with $m \ge 5$ dominate the spiral structure, as expected for a relatively thin disc where the transport is primarily local \citep{lr04,clc09}.  The $q=0.2$ disc shows more power in the lower-$m$ ($m = 2$--4) modes, similar to the behaviour seen in previous simulations \citep[e.g.,][]{lr05}.  We are thus satisfied that our isolated disc models are consistent with previous results, and that our numerical method is satisfactory.

\subsection{Disc with infall}

The central column in Fig.~\ref{fig:panels} shows the evolution of the disc with infall.  The behaviour is very different from that of either of the isolated discs.  The first notable difference is the formation of high-amplitude spiral density waves at $t \simeq 2000$, and the onset and growth of the GI occurs much faster than in the isolated discs.   In the presence of infall the disc shows much higher density contrasts (factors of $\simeq 2$--5) than the isolated discs, and well-ordered low-$m$ spiral structures. It is also worth noting that the disc does not fragment, despite being subject to high rates of infall; instead it is able to transport angular momentum fast enough to prevent any ``pile-up'' of the infalling material.

\begin{figure}
\centering
\resizebox{\hsize}{!}{
\includegraphics[angle=270]{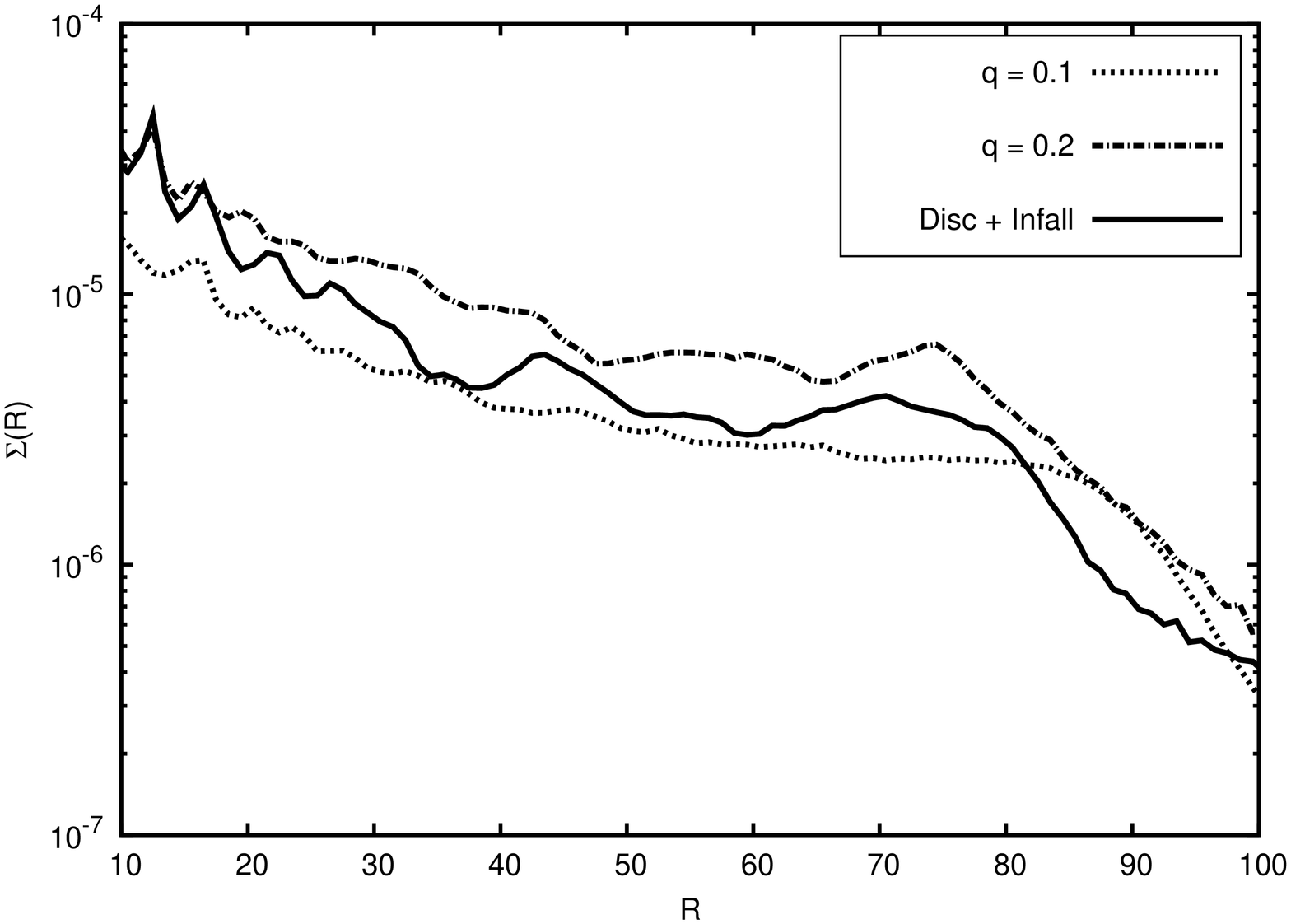}
}

\vspace*{6pt}

\resizebox{\hsize}{!}{
\includegraphics[angle=270]{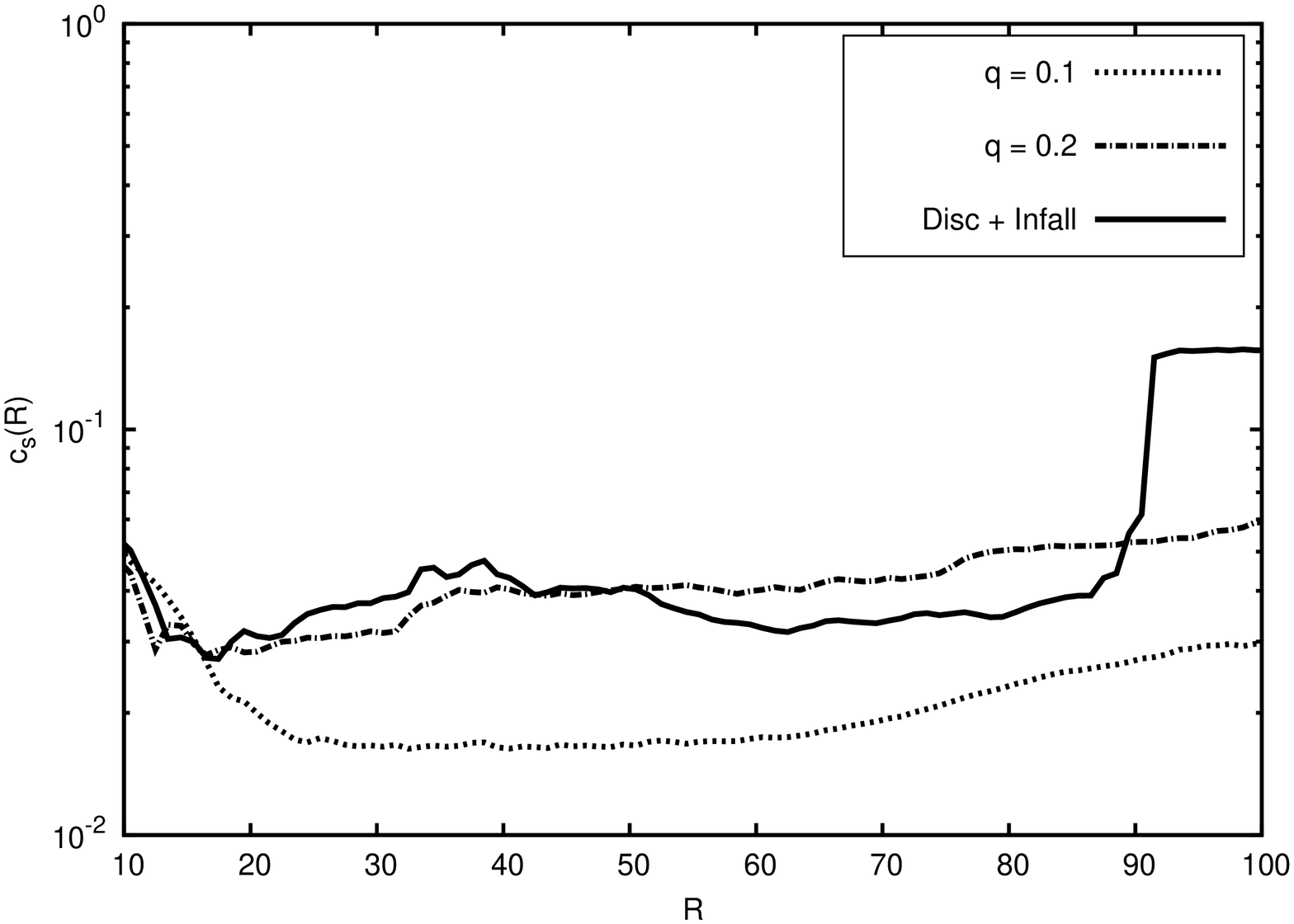}
}

\caption{Azimuthally-averaged radial profiles of surface density $\Sigma$ (top) and sound speed $c_s$ (bottom), plotted for all three simulations at $t=3500$.  In both cases the disc with infall lies between the $q=0.1$ and $q=0.2$ disc (except near the outer disc edge at $R=100$), suggesting that disc mass and temperature are not the primary differences between these simulations.}
\label{fig:panels_radprofile}
\end{figure}

\begin{figure}
\centering
\resizebox{\hsize}{!}{
\includegraphics[angle=270]{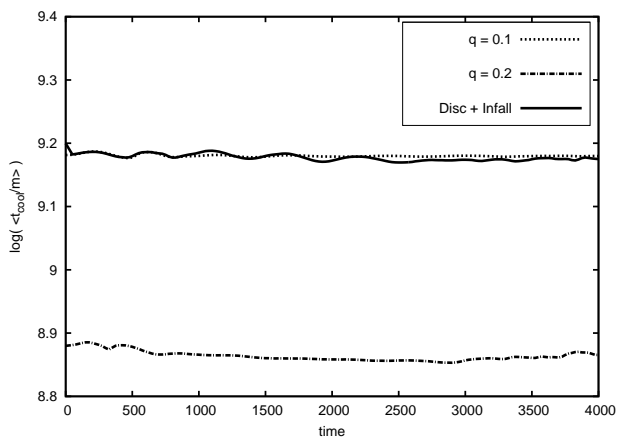}
}
\caption{Time evolution of the mass-weighted cooling time for the three different models, averaged over the radial range $R=60$--80, plotted in arbitrary units (on a logarithmic scale).  The infall model shows no significant differences with respect to the $q=0.2$ disc, and differs only by a factor of $\simeq 2$ from the $q=0.1$ disc, allowing us to rule out variations in the cooling time-scale as a factor in the enhanced GIs seen.}
\label{fig:tcool}
\end{figure}

In order to make a detailed comparison between the infall model and the isolated discs, it is first necessary to define the disc in the infall model (excluding envelope gas).  We define the disc as all gas within 3 scale-heights of the midplane, with the scale-height computed as $H = c_s/\Omega$ at the midplane.  Again we restrict our analysis to the region $20 \le R \le 80$, to prevent boundary effects from becoming dominant.  The total mass in the disc at $R \le 100$ at the end of the simulation is ($t = 5000$) is 0.14 M$_{\star}$, an increase of 0.04 M$_{\star}$ from the initial disc mass.  The disc is also more radially extended than the isolated discs, with significant mass at $R>100$ (and consequently higher temperatures at $R \gtrsim 90$).  Fig.\ref{fig:panels_radprofile} shows the azimuthally-averaged surface density and temperature (sound speed) profiles for all three models.  Within the region $20 \le R \le 80$ the surface density of the infall model lies between those of the two isolated discs.  At smaller radii the temperature of the infall model is similar to that of the $q=0.2$ disc, while at larger radii it lies between those of the $q=0.1$ and $q=0.2$ discs.  The fact that the surface density in the infall model is lower than in the $q=0.2$ disc rules out the $\simeq 40$\% increase in disc mass as being responsible for the changes seen in the infall model.  In addition, the mass-weighted cooling time-scales in the three models are very similar (see Fig.\ref{fig:tcool}), so we can rule out variations in the cooling time-scale as being responsible for difference between the disc-only simulations and the case with infall.  Instead, we find that the presence of an infalling envelope qualitatively changes the behaviour of the disc, and excites deep low-order spiral density waves.

As seen in Fig.\ref{fig:Qprof}, the infall model never reaches the self-regulated, $Q\simeq1$ state seen in simulations of isolated discs.  Instead, we see strong departures from $Q\simeq1$, which are primarily due to the substantial variations in the disc's surface density seen in Fig.\ref{fig:panels_radprofile}.  The Fourier analysis shows that the low-$m$ ($m=1$--4) modes dominate the spiral structures, and despite the lower surface density the power spectrum of the disc with infall shows no clear differences from the $q=0.2$ disc.  This suggests that infall can drive global transport of angular momentum even in relatively thin discs.

\begin{figure*}
\centering
\resizebox{\hsize}{!}{
\includegraphics[angle=0]{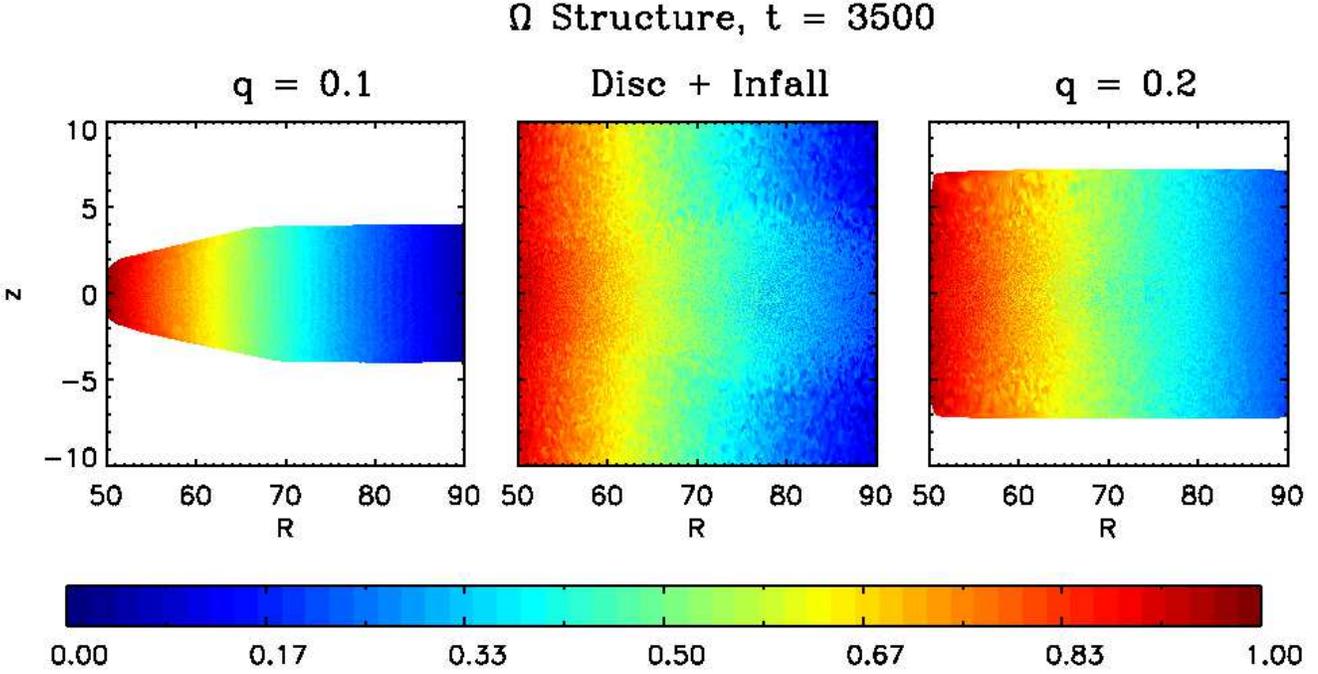}
}
\caption{Azimuthally-averaged $R-z$ projections of the orbital frequency $\Omega = v_\phi/R$ for the three different simulations, plotted at $t=3500$.  For clarity we have limited the plots to the radial range $R=50$--90, and for the disc-only models plotted only the region within within $\pm3H$ of the midplane.  In each case the values of $\Omega$ are normalised to the maximum value (that at $R=50$).  The vertical shear caused by the sub-Keplerian infall is clearly visible in the middle panel.}
\label{fig:panels_omega}
\end{figure*}

\begin{figure}
\centering
\resizebox{\hsize}{!}{
\includegraphics[angle=270]{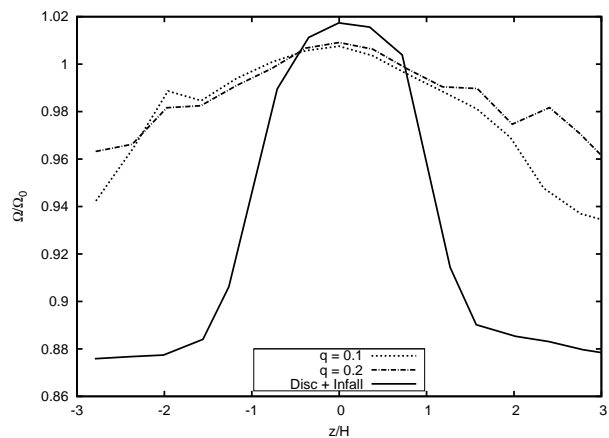}
}
\caption{The vertical profile of the orbital frequency $\Omega(z)$, azimuthally-averaged and normalised to the Keplerian orbital frequency $\Omega_{0}$.  The dotted and dashed lines denote the isolated discs ($q=0.1$ and $0.2$, respectively) while the solid line denotes the disc $+$ infall.  The upper layers of the disc in the infall model are strongly sub-Keplerian, due to the interaction with the (sub-Keplerian) infall. }
\label{fig:Omega}
\end{figure}

We can gain some insight into the behaviour of the disc subject to infall by looking at the vertical rotation profile.  Fig.\ref{fig:panels_omega} shows a 2-D, azimuthally-averaged projection of the orbital frequency in the discs (at $t=3500$), and Fig.~\ref{fig:Omega} shows the orbital frequency as a function of vertical position in the discs $\Omega(z)$, at $R = 75$ (effectively a vertical cross section of Fig.\ref{fig:panels_omega}).  In all three cases the midplane rotation is very close to Keplerian.  As expected the isolated discs show nearly constant $\Omega(z)$; the slight fall-off at high $z$ is primarily due to numerical effects, as the isolated disc models are not well-resolved for $|z| \gtrsim 2.5 H$ (where there is little mass, and therefore few SPH particles).  However, the model with infall shows strongly sub-Keplerian rotation away from the disc midplane: more than one scale-height away from the midplane, the rotation is sub-Keplerian by 5--10\%.  This occurs because infalling gas from the envelope is sub-Keplerian where it lands on the disc.  This vertical velocity shear has the potential to excite deeper spiral density waves than occur in the isolated discs, and drives the low-order spiral waves (which have a sub-Keplerian pattern speed).  

At this point it is instructive to consider the time-scales involved in both the GI and the vertical shearing.  The unstable modes of a gravitationally unstable disc grow on the dynamical time-scale, so if the vertical velocity shear is to play a significant role in modifying the behaviour of the GI it must occur on a similar (or shorter) time-scale.  We see from Fig.\ref{fig:Omega} that in the presence of infall the disc surface layer is sub-Keplerian by approximately 10\%.  The velocity difference across this shear is therefore $\simeq 0.1 \Omega R$, and the shearing time-scale $t_{sh} \sim H/(0.1 \Omega R) \sim (H/0.1R) t_{dyn}$.  In our disc $H/R \sim 0.1$, so the shearing time-scale is approximately equal to the dynamical time-scale.  This argument suggests that shearing does occur on a sufficiently short time-scale to influence the growth of GIs significantly, and supports our argument that the vertical velocity shear is responsible for the strong global modes seen in our disc in the presence of infall.  We note, however, that we cannot rule out the presence of other destabilising mechanisms also being present.

\begin{figure}
\centering
\resizebox{\hsize}{!}{
\includegraphics[angle=270]{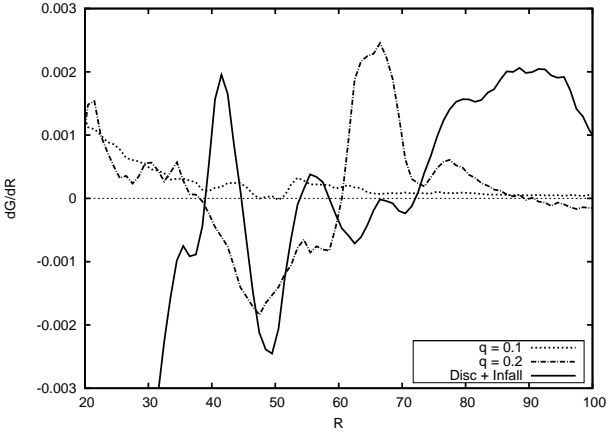}
}
\caption{Azimuthally-averaged profile of the gravitational torque in the discs, plotted at $t=3500$, with line-styles as in Fig.\ref{fig:Omega}.  The horizontal dotted line denotes $dG/dR=0$; negative torques correspond to the removal of angular momentum (accretion).  Unlike in the isolated discs, long-range gravitational torques dominate the angular momentum transport in the disc with infall.}
\label{fig:Torque}
\end{figure}

As mentioned above, we did not see any evidence for fragmentation with $\beta = 7.5$ even when the infall rate substantially exceeds the fragmentation threshold set by the local limit.  It appears, therefore, that when subject to infall the disc instead undergoes global transport of angular momentum, with consequent enhancement of accretion.  Unfortunately, although our simulations run for many dynamical periods their total duration is still relatively short compared to the (``viscous'') time-scale for angular momentum transport.  This makes determining the rate of angular momentum transport somewhat difficult, as at any given time in the simulations transients can be dominant. Moreover, any transport by low-$m$ spiral modes is intrinsically non-local \citep{bp99}, so looking purely at the local stresses \citep[as in][]{lr04} is not appropriate.  Instead, we  computed the differential gravitational torque $dG/dR$ as a function of radius, as this should highlight any non-local angular momentum transport.  The torque profiles from the three models are shown in Fig.~\ref{fig:Torque}.  In the $q=0.1$ disc the gravitational torques are small everywhere, with transport dominated by local stresses.  Substantially larger gravitational torques are seen in the $q=0.2$ disc, but the peaks in $dG/dR$ correspond to individual spiral density waves and ``cancel'' over relatively small radial scales, and when averaged over many orbits.  Moreover, the torques are negligible at large radii ($R \gtrsim 70$) in both isolated discs.  By contrast we see strong torques throughout the disc subject to infall, primarily negative at small radii ($R \lesssim 60$) and positive at larger radii, which persist over several orbital periods.  These torques are therefore transporting angular momentum outward through the disc on length-scales comparable to the disc radius.  If the torques on the disc were solely due to the interaction with the envelope we would expect to see $dG/dR < 0$ at all radii, as the infall is sub-Keplerian.  Instead we see both negative and positive torques at the midplane in different regions of the disc, which strongly suggests that global gravitational torques are the primary mechanism driving its evolution. 

\section{Discussion}\label{sec:dis}

\subsection{Limitations}
We have presented calculations on the effect of infall on to a gravitationally unstable accretion disc, but our approach is highly idealised.  This approach allows us to study the key physical processes in detail, but imposes some limitations when we apply our results to real astrophysical situations.

Our first major simplification is in the initial conditions of our infall model.  In order to ensure that we understand the various numerical effects in our calculations, we chose to let a rotating cloud fall on to an already-present disc, rather than letting the disc form self-consistently.  The advantages of our set-up are two-fold.  First, we can control the accretion rate on to the disc, and ensure that the ratio between infall rate and theoretical accretion rate is approximately constant over the duration of our simulation (as seen in Fig.\ref{fig:infallrate}).  Second, we can compare our results directly to our isolated disc models, where the physics is well-understood, and thus isolate the effects of infall from the myriad of other potential effects.  However, the trade-off is that the simulations are highly idealised, and not always realistic.  In particular, we note that quasi-spherical infall is only expected in the early stages of protostellar collapse, when the disc and envelope masses are likely to be much larger than those considered here \citep[e.g.,][]{boley09,vorobyov09}.  Our results have relevance to almost any case of infall on to a gravitationally unstable disc (as any infall is, by definition, sub-Keplerian), but we note that care be taken when applying our results to real systems.

Our second major simplification lies in our treatment of the disc thermodynamics.  Our scale-free cooling law has previously been studied in great detail \citep{g01,r03,lr04,clc09}, but it is recognised to be a poor approximation to real systems.  The effect of our scale-free cooling law can be seen in Fig.~\ref{fig:panels_radprofile}: the disc temperature increases slightly with radius.  Real discs almost invariably have cooling time-scales that are shorter (relative to the local dynamical time-scale) at large radii than at small radii, and are thus not scale-free \citep[e.g.,][]{rafikov07,clarke09}.  A more realistic treatment would require an opacity-based cooling prescription \citep[e.g.,][]{boley06,s07}, but this would introduce several new free parameters to the problem.  Again, our simplifications are not entirely physical in this regard, but do allow us to study the important processes in detail.  Essentially we have chosen to perform a well-controlled numerical experiment instead of a physically realistic simulation, and our results should be interpreted with this in mind.

\subsection{Comparison to Previous Work}
The majority of previous work in this area has studied the transport properties of isolated gravitationally unstable discs \citep{lb94,lr04,rla05,boley06,clc09}.  Our reference simulations exhibit the same behaviour as in these previous studies, as discussed in Section \ref{sec:disc_only}.  However, the influence of infall on the evolution of GIs has not yet been explored in great detail.  \citet{krumholz07} studied angular momentum transport in a self-consistently-formed disc subject to a very high rate of infall.  They found very high accretion rates, equivalent to $\sim$30\% of the total disc mass per dynamical time-scale, with effective $\alpha$-values that exceeded unity.  Most of the power was found in the $m=1$ mode, and \citet{krumholz07} attributed this very rapid accretion to the SLING instability \citep{adams89,shu90}.  We note, however, that the discs formed in these simulations were much more massive than those considered here, with $q \sim 0.5$--1.  It has long been known that low-order spiral modes can drive rapid accretion in massive discs \citep[e.g.,][]{lka98}, and in this regards the results of \citet{krumholz07} are not directly comparable to those of our simulations.

In addition, a number of recent studies have used one- and two-dimensional simulations to study the formation and evolution of protostellar discs \citep[e.g.,][]{hueso05,vb07,vb09,vorobyov09,vd10,zhu09,zhu10}.  These simulations are less computationally intensive than 3-D simulations, and are therefore able to follow the evolution of the system for much longer time-scales.  They also make use of more physically realistic prescriptions for both infall and thermodynamics, forming discs self-consistently from collapsing clouds and incorporating realistic models for radiative heating and cooling.  These simulations generally predict that most of the GIs' power is found in low-order spiral modes, due to both infall and the relatively high masses of the discs which form.  In addition, many of these simulations have been seen to exhibit transient accretion outbursts, triggered in some cases by the accretion of bound clumps of gas \citep[e.g.,][]{vb06,vb07} and in others by interaction between gravitational instability in the outer disc and layered accretion in the inner disc \citep[e.g.,][]{zhu10}.  Unfortunately it is not straightforward to draw direct comparisons between these results and ours, due to the complex effects of both the cooling and infall prescriptions used.  We note, however, that the vertical shear effect observed in our simulations is intrinsically a three-dimensional phenomenon, and consequently cannot be observed in 2-D, vertically-integrated simulations.  Our results point towards an additional mechanism for driving transient transport of angular momentum, and lend further weight to the well-established idea that low-order spiral modes drive accretion in protostellar discs.

By contrast, to date only a handful of similar studies have been conducted in three dimensions.  Most relevant here is the work of \citet{boley09} and \citet{kratter10}, who used three-dimensional hydrodynamics to study the formation and evolution of protostellar discs.  \citet{boley09} used grid-based hydrodynamics with a similar set-up to that considered here: prescribed infall on to an already-present disc.  In some cases the discs fragmented, while in others angular momentum transport was dominated by low-order spiral density waves.  We note, however, that the discs in the simulations of \citet{boley09} are significantly more massive than ours ($q \sim 0.3$--0.5), increasing the importance of global modes.  Given the additional differences between the simulations (most notably in the adopted cooling models) it is difficult to make direct comparisons, but in general our result -- that infall on to the disc enhances the importance of global modes -- seems consistent with those of \citet{boley09}.

By contrast, in the models of \citet{kratter10} discs form and evolve in a self-consistent manner, and they were able to explore a larger range in parameter space than we have achieved here.  They found, as we do, that low-order spiral modes dominate the transport of angular momentum, although this again may in part be driven by the fact that their discs are somewhat more massive than ours.  However, \citet{kratter10} found that infall rates of $\gtrsim3$ times the disc accretion rate typically led to fragmentation, while we find that no fragmentation despite an infall rate nearly an order of magnitude higher than the ``local limit'' for disc accretion.  Unfortunately it is not straightforward to compare these apparently contradictory results directly, due to the different prescriptions used for disc thermodynamics.  \citet{kratter10} adopted an isothermal equation of state, and defined their models with two parameters (representing the accretion rate and angular momentum of the infalling gas).  By contrast, we adopt an adiabatic equation of state with a parametrized cooling function, with the spherical envelope given a uniform initial temperature.  The prescribed cooling time-scale is much longer than the dynamical time-scale (by a factor $\beta = 7.5$), so the infalling gas is effectively adiabatic.  This results in slight heating of the infalling gas, and a corresponding increase in temperature in the disc.  The increase in the disc temperature is not dramatic (see Fig.\ref{fig:panels_radprofile}), but given that the rate of spherical accretion scales as $c_s^3$ even this small difference could account for the factor of $\sim 3$ discrepancy between our results and those of \citet{kratter10}.  Additional simulations, using different initial cloud temperatures and cooling laws, are required to investigate this issue in more detail, but such simulations are beyond the scope of this paper.  It is not clear whether the isothermal or adiabatic approximation is more relevant to real discs; most probably both have some validity in different regions of the disc.  We thus regard our results as complementary to those of \citet{kratter10}, and encourage further work in this area.

\subsection{Applications to Observed Systems}
Our results have obvious applications to the physics of star formation, in particular the formation of low-mass ($\sim 1$M$_{\odot}$) stars.  Observations suggest that essentially all low-mass stars form with discs \citep[e.g.,][]{hll01}, and disc accretion is thought to play a major role in the build-up of stellar mass.  Moreover, in the earliest, embedded phases gravitational instability is likely to be the dominant mechanism for angular momentum transport: such discs are insufficiently ionized to sustain transport via magnetohydrodyamic turbulence \citep{mt95,g96}, but both observations and theory suggest that they are indeed massive enough to be gravitationally unstable \citep[e.g.,][]{greaves08,andrews09,hueso05,vorobyov09}.  Our results argue that accretion in such discs is likely to be highly transient, and in general terms are consistent with a picture where the bulk of the stellar mass is accreted during a small number of intense outbursts \citep[e.g.,][]{armitage01,lr05,vb06,vb10,zhu10}.

The consequences of our results for massive star formation are less clear.  Although discs are expected to form around massive, forming stars, observational evidence of their existence is somewhat thin \citep[e.g.,][]{cesaroni06,cesaroni07}.  It is clear, however, that if such discs do indeed exist they are likely to be gravitationally unstable, but in this scenario the maximum stable disc accretion rate ($\lesssim 10^{-5}$M$_{\odot}$yr${^{-1}}$; \citealt{levin03,cesaroni06,rafikov07}) is much too low for these massive stars to accumulate their mass in a plausible time-scale.  It has previously been suggested that the formation of massive stars is likely to be dominated by transient episodes and highly variable accretion \citep{cesaroni07}.  Our results suggest that the global gravitational torques driven by infall on to the disc result in exactly this type of behaviour, and may be an important accretion mechanism in massive star formation.

In addition, we suggest that our results may have important consequences for the formation of massive stars close to super-massive black holes.  A large population of massive O- and Wolf-Rayet-type stars is now known to exist within $\sim 0.1$pc of the super-massive black hole (SMBH) at the centre of the Galaxy \citep{genzel03,ghez05}, and a popular scenario for the formation of these stars is ``in situ'' formation via the fragmentation of an accretion disc around the SMBH \citep{lb03,nayakshin06}.  This picture has a number of attractive features, but an open question has always been how the stars attain their final masses.  Both analytic theory and numerical simulations suggest that the initial fragment masses are small, $\sim 1$M$_{\odot}$, and that the bulk of the stellar mass is subsequently accreted from the SMBH disc \citep[e.g.,][]{rda08,br08}.  However, the estimated infall rates through the Hill sphere on to these protostellar discs are extremely high, $\sim 10^{-4}$M$_{\odot}$yr$^{-1}$ \citep{ml04}, and in the local limit these discs are expected to fragment, preventing rapid growth of the protostars and limiting the resulting stellar masses \citep[see also][]{ml05,km06}.  Our results suggest that discs subject to high infall rates may instead be able to transport angular momentum much more rapidly, through global modes of the GI, and this mechanism provides a potential solution to the ``accretion problem'' of massive star formation at the centre of the Galaxy.


\section{Summary}\label{sec:conc}
We have presented numerical simulations of gravitationally unstable accretion discs subject to infall from a surrounding envelope.  Our numerical set-up is highly idealised, but this allows us to study the angular momentum transport properties of the system in detail.  Our disc has a relatively slow cooling rate ($t_{\mathrm {cool}} \Omega = 7.5$), and we find that the disc does not fragment even though the infall rate on to the disc is an order of magnitude greater than the ``quasi-steady'' accretion rate in the self-regulating self-gravitating disc.  Instead, despite relatively low disc masses ($M_d/M_\star \simeq 0.14$), we see evidence that angular momentum is transported rapidly by torques from low-order, global, spiral density waves, which are excited by the interaction between the disc and the infalling envelope.  This drives accretion at a rate significantly higher than is possible in a local model, and we suggest that this mechanism may play an important role in a number of different astrophysical systems.

\section*{Acknowledgments}
We are grateful for stimulating discussions with a number of colleagues, in particular Giuseppe Lodato, Phil Armitage, Peter Cossins, Chris Matzner, Kaitlin Kratter \& Steve Balbus.  We also thank the anonymous referee for several useful comments.  DH is supported by an Oort Scholarship from Leiden Observatory, and by a Huygens Scholarship from the Netherlands Organization for International Co-operation in Higher Education (NUFFIC).  RDA \& YL acknowledge support from the from the Netherlands Organisation for Scientific Research (NWO) through VIDI grant 639.042.607.  RDA also acknowledges further support from the NWO through VIDI grant 639.042.404, and from a Science \& Technology Facilities Council (STFC) Advanced Fellowship (ST/G00711X/1).  Visualizations of SPH simulations were created using \emph{SPLASH} \citep{price07}.  The simulations presented in this paper were run on the {\it Huygens} supercomputer, supported by Netherlands National Computing Facility Project SH-080-08.



\begin{thebibliography}{99}
\bibitem[\protect\citeauthoryear{Adams, Ruden, \& Shu}{Adams et al.}{1989}]{adams89} Adams, F.~C., Ruden, S.~P., Shu, F.~H., 1989, ApJ, 347, 959 
\bibitem[\protect\citeauthoryear{Andrews \& Williams}{2005}]{aw05} Andrews, S.~M., Williams, J.~P., 2005. ApJ 631, 1134
\bibitem[\protect\citeauthoryear{Andrews et~al.}{2009}]{andrews09} Andrews, S.~M., Wilner, D.~J., Hughes ,A.~M., Qi, C., Dullemond, C.~P., 2009, ApJ, 700, 1502 
\bibitem[\protect\citeauthoryear{Alexander et~al.}{2008}]{rda08} Alexander, R.~ D., Armitage, P.~J., Cuadra, J., Begelman, M.~C., 2008, ApJ, 674, 927
\bibitem[\protect\citeauthoryear{Armitage, Livio \& Pringle}{Armitage et al.}{2001}]{armitage01} Armitage, P,~J., Livio, M., Pringle, J.~E., 2001, MNRAS, 324, 705
\bibitem[\protect\citeauthoryear{Balsara}{1995}]{b95} Balsara, D.~S., 1995, J.~Comput.~Phys., 121, 357
\bibitem[\protect\citeauthoryear{Balbus \& Hawley}{1991}]{bh91} Balbus, S.~A., Hawley, J.~F., 1991, ApJ, 376, 214
\bibitem[\protect\citeauthoryear{Balbus \& Hawley}{1998}]{bh98} Balbus S.~A., Hawley J.~F., 1998, RvMP, 70, 1 
\bibitem[\protect\citeauthoryear{Balbus \& Papaloizou}{1999}]{bp99} Balbus, S.~A., Papaloizou, J.~C.~B., 1999, ApJ, 521, 650
\bibitem[\protect\citeauthoryear{Bell \& Lin}{1994}]{bl94} Bell, K.R., Lin, D.N.C., 1994, ApJ, 427, 987
\bibitem[\protect\citeauthoryear{Bell et~al.}{1997}]{bell97} Bell, K.R., Cassen, P.M., Klahr, H.H., Henning, Th., 1997, ApJ, 486, 372
\bibitem[\protect\citeauthoryear{Boley et~al.}{2006}]{boley06} Boley A.~C., Mej{\'{\i}}a A.~C., Durisen R.~H., Cai K., Pickett M.~K., D'Alessio P., 2006, ApJ, 651, 517
\bibitem[\protect\citeauthoryear{Boley et~al.}{2007}]{boley07} Boley, A.C., Hartquist, T.W., Durisen, R.H., Michael, S., 2007, ApJ, 656, 89
\bibitem[\protect\citeauthoryear{Boley}{2009}]{boley09} Boley, A.~C., 2009, ApJ, 695, L53 
\bibitem[\protect\citeauthoryear{Bonnell \& Rice}{2008}]{br08} Bonnell I.~A., Rice W.~K.~M., 2008, Sci, 321, 1060 
\bibitem[\protect\citeauthoryear{Calvet, Hartmann \& Strom}{Calvet et~al.}{2000}]{calvet_ppiv} Calvet, N., Hartmann, L., Strom, S.~E., 2000, in Mannings, V., Boss. A.P., Russell, S.S., eds, {\it Protostars \& Planets IV}, U.~Arizona Press, Tuscon, 377 
\bibitem[\protect\citeauthoryear{Cesaroni et~al.}{2006}]{cesaroni06} Cesaroni, R., Galli, D., Lodato, G., Walmsley, M., Zhang, Q., 2006, Natur, 444, 703 
\bibitem[\protect\citeauthoryear{Cesaroni et~al.}{2007}]{cesaroni07} Cesaroni, R., Galli, D., Lodato, G., Walmsley, C.~M., Zhang, Q., 2007, in Reipurth, B., Jewitt, D., Keil, K., eds, Protostars \& Planets V, Univ. Arizona Press, Tuscon, 197
\bibitem[\protect\citeauthoryear{Clarke}{2009}]{clarke09} Clarke C.~J., 2009, MNRAS, 396, 1066 
\bibitem[\protect\citeauthoryear{Cossins, Lodato \& Clarke}{Cossins et~al.}{2009}]{clc09} Cossins, P., Lodato, G., Clarke, C.~J., 2009, MNRAS, 393, 1157
\bibitem[\protect\citeauthoryear{Cuadra et~al.}{2006}]{cuadra06} Cuadra, J., Nayakshin, S., Springel, V., di Mattero, T., 2006, MNRAS, 366, 358
\bibitem[\protect\citeauthoryear{Durisen et~al.}{2007}]{d07} Durisen, R.~H., Boss, A.~P., Mayer, L., Nelson, A.~F., Quinn, T., Rice, W.~K.~M., 2007, in Reipurth, B., Jewitt, D., Keil, K., eds, Protostars \& Planets V, Univ. Arizona Press, Tuscon, 607
\bibitem[\protect\citeauthoryear{Gammie}{1996}]{g96} Gammie C.~F., 1996, ApJ, 457, 355
\bibitem[\protect\citeauthoryear{Gammie}{2001}]{g01} Gammie C.~F., 2001, ApJ, 553, 174
\bibitem[\protect\citeauthoryear{Genzel et~al.}{2003}]{genzel03} Genzel R., et~al., 2003, ApJ, 594, 812 
\bibitem[\protect\citeauthoryear{Ghez et~al.}{2005}]{ghez05} Ghez A.~M., Salim S., Hornstein S.~D., Tanner A., Lu J.~R., Morris M., Becklin E.~E., Duch{\^e}ne G., 2005, ApJ, 620, 744 
\bibitem[\protect\citeauthoryear{Greaves et~al.}{2008}]{greaves08} Greaves, J.~S., Richards, A.~M.~S., Rice, W.~K.~M., Muxlow, T.~W.~B., 2008, MNRAS, 391, L74 
\bibitem[\protect\citeauthoryear{Haisch, Lada, \& Lada}{Haisch et~al.}{2001}]{hll01} Haisch K.~E., Jr., Lada E.~A., Lada C.~J., 2001, ApJ, 553, L153 
\bibitem[\protect\citeauthoryear{Hueso \& Guillot}{2005}]{hueso05} Hueso R., Guillot T., 2005, A\&A, 442, 703
\bibitem[\protect\citeauthoryear{Kenyon et~al.}{1990}]{kenyon90} Kenyon, S.~J., Hartmann, L.~W., Strom, K.~M., Strom, S.~E., 1990, AJ, 99, 869 
\bibitem[\protect\citeauthoryear{King, Pringle \& Livio}{King et~al.}{2007}]{kpl07} King, A.~R., Pringle, J.~E., Livio, M., 2007, MNRAS, 376, 1740
\bibitem[\protect\citeauthoryear{Kratter \& Matzner}{2006}]{km06} Kratter, K.~M., Matzner, C.~D., 2006, MNRAS, 373, 1563 
\bibitem[\protect\citeauthoryear{Kratter et~al.}{2010}]{kratter10} Kratter, K.~M., Matzner, C.~D., Krumholz, M.~R., Klein, R.~I., 2010, ApJ, 708, 1585
\bibitem[\protect\citeauthoryear{Krumholz, Klein \& McKee}{Krumholz et~al.}{2007}]{krumholz07} Krumholz, M.~R., Klein, R.~I., McKee, C.~F., 2007, ApJ, 656, 959
\bibitem[\protect\citeauthoryear{Laughlin \& Bodenheimer}{1994}]{lb94} Laughlin, G., Bodenheimer, P., 1994, ApJ, 325, 231 
\bibitem[\protect\citeauthoryear{Laughlin, Korchagin, \& Adams}{Laughlin et~al.}{1998}]{lka98} Laughlin G., Korchagin V., Adams F.~C., 1998, ApJ, 504, 945 
\bibitem[\protect\citeauthoryear{Levin}{2003}]{levin03} Levin, Y., 2003, preprint(astro-ph/0307084)
\bibitem[\protect\citeauthoryear{Levin \& Beloborodov}{2003}]{lb03} Levin, Y., Beloborodov, A.~M., 2003, ApJ, 590, 33
\bibitem[\protect\citeauthoryear{Levin}{2007}]{levin07} Levin, Y., 2007, MNRAS, 374, 515
\bibitem[\protect\citeauthoryear{Lodato \& Rice}{2004}]{lr04} Lodato, G., Rice, W.~K.~M., 2004, MNRAS, 351, 630
\bibitem[\protect\citeauthoryear{Lodato \& Rice}{2005}]{lr05} Lodato, G., Rice, W.~K.~M., 2005, MNRAS, 358, 1489
\bibitem[\protect\citeauthoryear{Lodato}{2008}]{lodato08} Lodato, G., 2008, New Astronomy Reviews, 52, 21
\bibitem[\protect\citeauthoryear{Lynden-Bell \& Kalnajs}{1972}]{lbk72} Lynden-Bell D., Kalnajs A.~J., 1972, MNRAS, 157, 1 
\bibitem[\protect\citeauthoryear{Matsumoto \& Tajima}{1995}]{mt95} Matsumoto, R., Tajima, T., 1995, ApJ, 445, 767 
\bibitem[\protect\citeauthoryear{Matzner \& Levin}{2005}]{ml05} Matzner, C.~D., Levin, Y., 2005, ApJ, 628, 81
\bibitem[\protect\citeauthoryear{Mej{\'{\i}}a et~al.}{2005}]{mejia05} Mej{\'{\i}}a, A.~C., Durisen, R.~H., Pickett, M.~K., Cai, K., 2005, ApJ, 619, 1098 
\bibitem[\protect\citeauthoryear{Milosavljevi{\'c} \& Loeb}{2004}]{ml04} Milosavljevi{\'c} M., Loeb A., 2004, ApJ, 604, L45 
\bibitem[\protect\citeauthoryear{Monaghan \& Gingold}{1983}]{mg83} Monaghan, J.~J., Gingold, R.~A., 1983, J.~Comput.~Phys., 52, 374
\bibitem[\protect\citeauthoryear{Murray}{1996}]{m96} Murray, J.~R., 1996, MNRAS, 279, 402 
\bibitem[\protect\citeauthoryear{Nayakshin}{2006}]{nayakshin06} Nayakshin, S., 2006, MNRAS, 372, 143
\bibitem[\protect\citeauthoryear{Nelson}{2006}]{nelson06} Nelson, A.~F., 2006, MNRAS, 366, 358
\bibitem[\protect\citeauthoryear{Price}{2007}]{price07} Price, D.~J., 2007, PASA, 24, 159
\bibitem[\protect\citeauthoryear{Rafikov}{2007}]{rafikov07} Rafikov, R.~R., 2007, ApJ, 662, 642
\bibitem[\protect\citeauthoryear{Rafikov}{2009}]{rafikov09} Rafikov, R.~R., 2009, ApJ, 704, 281 
\bibitem[\protect\citeauthoryear{Rice et~al.}{2003}]{r03} Rice W.~K.~M., Armitage P.~J., Bate M.~R., Bonnell I.~A., 2003, MNRAS, 339, 1025
\bibitem[\protect\citeauthoryear{Rice, Lodato, \& Armitage}{Rice et~al.}{2005}]{rla05} Rice W.~K.~M., Lodato G., Armitage P.~J., 2005, MNRAS, 364, L56
\bibitem[\protect\citeauthoryear{Shakura \& Sunyaev}{1973}]{ss73} Shakura, N.~I., Sunyaev, R.~A., 1973, A\&A, 24, 337
\bibitem[\protect\citeauthoryear{Shu et al.}{1990}]{shu90} Shu, F.~H., Tremaine, S., Adams, F.~C., Ruden, S.~P., 1990, ApJ, 358, 495 
\bibitem[\protect\citeauthoryear{Springel}{2005}]{springel05} Springel, V., 2005, MNRAS, 364, 1105
\bibitem[\protect\citeauthoryear{Stamatellos et~al.}{2007}]{s07} Stamatellos D., Whitworth A.~P., Bisbas T., Goodwin S., 2007, A\&A, 475, 37 
\bibitem[\protect\citeauthoryear{Toomre}{1964}]{toomre64} Toomre, A., 1964, ApJ, 139, 1217
\bibitem[\protect\citeauthoryear{Visser \& Dullemond}{2010}]{vd10} Visser, R., Dullemond, C.~P., 2010, A\&A, 519, A28 
\bibitem[\protect\citeauthoryear{Vorobyov \& Basu}{2006}]{vb06} Vorobyov, E.~I., Basu, S., 2006, ApJ, 650,956
\bibitem[\protect\citeauthoryear{Vorobyov \& Basu}{2007}]{vb07} Vorobyov, E.~I., Basu, S., 2007, MNRAS, 381, 1009
\bibitem[\protect\citeauthoryear{Vorobyov}{2009}]{vorobyov09} Vorobyov, E.~I., 2009, ApJ, 704, 715
\bibitem[\protect\citeauthoryear{Vorobyov \& Basu}{2009}]{vb09} Vorobyov, E.~I., Basu, S., 2009, MNRAS, 393, 822
\bibitem[\protect\citeauthoryear{Vorobyov \& Basu}{2010}]{vb10} Vorobyov, E.~I., Basu, S., 2010, ApJ, 719, 1896
\bibitem[\protect\citeauthoryear{Zhu et al.}{2009}]{zhu09} Zhu, Z., Hartmann, L., Gammie, C., McKinney, J.~C., 2009, ApJ, 701, 620 
\bibitem[\protect\citeauthoryear{Zhu, Hartmann, \& Gammie}{Zhu et al.}{2010}]{zhu10} Zhu, Z., Hartmann, L., Gammie, C., 2010, ApJ, 713, 1143 
\label{lastpage}
\end{thebibliography}
\end{document}